\documentclass[10pt]{iopart}

\expandafter\let\csname equation*\endcsname\relax
\expandafter\let\csname endequation*\endcsname\relax

\usepackage{bm}					
\usepackage{amsmath}
\usepackage{setspace}
\usepackage{algorithm}
\usepackage{algorithmic}
\usepackage{mathrsfs}  
\usepackage{graphicx}
\usepackage{hyperref} 
\usepackage{booktabs}
\usepackage{tabularx}
\usepackage{mathtools}
\usepackage{amssymb}
\usepackage{physics}
\usepackage{lipsum}  
\usepackage{algorithm}
\usepackage{comment}
\usepackage{orcidlink}
\usepackage{amsfonts}

\allowdisplaybreaks				
	
	\usepackage{float}				
		\usepackage{graphicx}			

\begin{document}	
\title{Financial Fraud Detection using Quantum Graph Neural Networks}
			
\author{Nouhaila Innan\orcidlink{0000-0002-1014-3457}$^{1,2}$\footnote{nouhaila.innan-etu@etu.univh2c.ma}, Abhishek Sawaika$^3$, Ashim Dhor$^4$, Siddhant Dutta$^5$, Sairupa Thota$^6$, Husayn Gokal$^7$, Nandan Patel$^8$, Muhammad Al-Zafar Khan\orcidlink{0000-0002-1147-7782}$^{2,9}$\footnote{muhammadalzafark@gmail.com}, Ioannis Theodonis$^{10}$ and Mohamed Bennai$^{1}$}
\address{$^1$ Quantum Physics and Magnetism Team, LPMC, Faculty of Sciences Ben M'sick, Hassan II University of Casablanca, Morocco}
\address{$^2$ Quantum Formalism Fellow, Zaiku Group Ltd, Liverpool, United Kingdom}
\address{$^3$ Indian Institute of Technology, Bombay, India}
\address{$^4$ Department of Data Science and Engineering, Indian Institute of Science Education and Research Bhopal, India}
\address{$^5$ Dwarkadas J. Sanghvi College of Engineering}
\address{$^6$ Guru Nanak Institutions Technical Campus, Hyderabad, India}
\address{$^7$ ASTI Academy, Dubai, UAE}
\address{$^8$ Vellore Institute of Technology, Chennai, India}
\address{$^9$ Robotics, Autonomous Intelligence, and Learning Laboratory (RAIL), School of Computer Science and Applied Mathematics, University of the Witwatersrand, 1 Jan Smuts Ave, Braamfontein, Johannesburg 2000, Gauteng, South Africa}
\address{$^{10}$ Department of Physics, School of Applied Mathematical and Physical Sciences, National Technical University of Athens, Greece.}

\vspace{10pt}
\begin{indented}
\item[]\date{\today}
\end{indented}

\begin{abstract}
Financial fraud detection is essential for preventing significant financial losses and maintaining the reputation of financial institutions. However,  conventional methods of detecting financial fraud have limited effectiveness, necessitating the need for new approaches to improve detection rates.
In this paper, we propose a novel approach for detecting financial fraud using Quantum Graph Neural Networks (QGNNs). QGNNs are a type of neural network that can process graph-structured data and leverage the power of Quantum Computing (QC) to perform computations more efficiently than classical neural networks. Our approach uses Variational Quantum Circuits (VQC) to enhance the performance of the QGNN. In order to evaluate the efficiency of our proposed method, we compared the performance of QGNNs to Classical Graph Neural Networks using a real-world financial fraud detection dataset. The results of our experiments showed that QGNNs achieved an AUC of $0.85$, which outperformed classical GNNs. Our research highlights the potential of QGNNs and suggests that QGNNs are a promising new approach for improving financial fraud detection.
		
\end{abstract}
\vspace{2pc}
\ioptwocol

\section{\label{sec:level1}Introduction}

In today's dynamic technological landscape, the rapid surge in digital transactions and financial activities has ushered in unprecedented conveniences. However, this digital transformation has also fostered a parallel escalation in financial fraud, posing significant threats to global industries and economies. As financial operations increasingly adopt digital complexities, the susceptibility to exploitation from malicious actors increases, leading to the pursuit of innovative fraud detection methods \cite{AS01}.

The importance of fraud detection extends beyond a single industry, resonating across diverse sectors, including Banking, e-Commerce, Healthcare, FinTech, and Insurance. The integrity of financial transactions is pivotal for optimal functionality in these domains, making robust fraud detection systems indispensable. The banking sector, exemplifying the forefront of this challenge, confronts expanding attack surfaces due to rising online and mobile banking adoption \cite{AS04}. Conventional rule-based approaches, while partially effective, struggle to match the ever-evolving tactics of fraudsters. This necessitates advanced, adaptable techniques in order to keep pace with the dynamic landscape of financial fraud \cite{AS05}.

The research community is constantly exploring new ways to detect fraud, as the methods used by fraudsters become increasingly sophisticated. An emerging avenue combines Quantum Machine Learning (QML), and Graph Neural Networks (GNNs); a groundbreaking approach poised to revolutionise financial fraud detection \cite{AS08}. By harnessing the distinct attributes of Quantum
Computing (QC) –- superposition and entanglement –- and integrating them with the effectiveness of GNNs, this hybrid paradigm endeavours to tackle the intricate and ever-changing realm of financial fraud. In this paper, we investigate the challenges of financial fraud detection and the potential of Quantum Graph Neural Networks (QGNNs) as a transformative approach to address these challenges. We conduct comprehensive analyses and experiments to demonstrate the efficacy and viability of QGNNs for fraud detection. Our findings contribute to the discourse on advanced fraud detection techniques and pave the way for enhanced security in the financial ecosystem.

 QML emerges at the intersection of QC and conventional Machine Learning (ML) techniques, offering promising avenues for solving complex problems in various domains. In the context of financial fraud detection, QML presents an intriguing approach that harnesses the inherent power of quantum computations to address the intricate challenges posed by fraudulent activities within the financial sector. By leveraging QGNNs, the combination of quantum and graphical techniques offers a novel framework for analysing and classifying financial transactions.

In financial fraud detection, QML encodes transactional data into quantum states. This can improve the efficiency and accuracy of fraud detection by enabling efficient parallel processing and enhanced feature representation. The unique property of quantum entanglement facilitates the capture of intricate relationships among transactional features. This fosters a more comprehensive understanding of fraudulent patterns.

However, this emerging paradigm also poses significant challenges, including the interpretability of quantum computations and concerns over quantum system security. Nonetheless, the promising outcomes showcased in this study highlight the potential of QML to significantly impact fraud detection by improving accuracy and scalability, opening the door to a new era of enhanced financial security. The inherent ability of quantum systems to process and model complex relationships among variables aligns well with the intricacies of financial transactions.

In the realm of fraud detection, conventional methods often struggle to navigate the intricate tapestry of relationships and intricate patterns that underlie fraudulent activities. As the complexity and diversity of fraudulent schemes continue to evolve, the demand for adaptable and nuanced detection techniques becomes increasingly apparent. In response, graph-based methods have emerged as a promising avenue, harnessing the intrinsic strength of graph structures to capture and dissect the multifaceted interconnections that define fraudulent behaviours.

Built upon the foundation of the mathematical study of graph theory, these methods provide an intuitive and comprehensive framework to model, depict, and analyse the labyrinthine-like relationships among entities. From individuals and accounts to transactions and organisations, a broad spectrum of entities can be seamlessly encapsulated within interconnected nodes and edges. This unique arrangement empowers graph structures to illustrate the intricate dynamics inherent in fraudulent schemes vividly. Diverging from conventional approaches, which rely on isolated data points and predetermined rules, graph-based methods can unveil concealed connections and dependencies, thereby unveiling anomalous or suspicious behaviours that might go unnoticed \cite{AS02}.

The true effectiveness of graph structures in fraud detection lies in their unparalleled ability to uncover latent patterns that elude customary detection techniques. Fraudulent endeavours rarely occur in isolation; rather, they unravel through intertwined actions involving multiple entities. These connections, often obscured by noise and obfuscation, are revealed through meticulous graph analysis. By representing and exploring these relationships, graph-based methods empower fraud detection systems to transcend the confines of isolated data points, capturing the holistic context in which fraudulent activities manifest and thrive.

This quantum potential extends seamlessly to the detection of financial fraud, where quantum-inspired techniques can transform the landscape. By integrating the strengths of quantum mechanics, QC could potentially revolutionise classical ML algorithms used for fraud detection. Techniques such as Quantum Neural Networks (QNNs) can harness the property of quantum parallelism to process and analyse financial data more efficiently. Exploiting superposition, these networks can simultaneously examine multiple data points, accelerating tasks like pattern recognition and data classification, crucial elements in detecting fraudulent activities. Furthermore, quantum feature mapping enhances the ability to identify intricate patterns in financial transactions that might evade classical methods. This intersection of QML and financial fraud detection promises a future where quantum-inspired techniques amplify the accuracy and efficiency of identifying and combating fraudulent behaviours.

The integration of quantum properties into graph-based fraud detection methodologies is motivated by the imperative to confront the escalating sophistication of fraudulent activities and the inherent limitations of classical computing paradigms. In today's landscape of intricate financial systems, fraudulent behaviours manifest intricate and interwoven patterns that defy conventional computational approaches. These conventional methods struggle to unravel the complexities inherent in fraud networks, marked by exponential data growth and non-linear dynamics. As a result, a critical need arises for innovative techniques that transcend classical computing boundaries, leading us to explore the distinctive attributes of QC.

While graph-based fraud detection methods have shown their ability to capture the intricate relationships within fraudulent activities, they face limitations in handling the explosive data growth and complex dynamics of fraud networks \cite{AS07}. The conventional algorithms grounded in classical computing struggle to navigate the intricate patterns that define these networks, often characterised by non-linear structures and extensive solution spaces. This disparity underscores a compelling necessity for pioneering methodologies capable of transcending the confines of classical computation. By delving into quantum properties, we seek to bridge this gap, illuminating concealed fraud schemes and redefining the boundaries of fraud detection in the context of modern financial systems \cite{AS09}.

The primary objective of this study is to conduct a comprehensive investigation into the effectiveness of QGNNs for financial fraud detection. Specifically, we aim to comprehensively assess and analyse the potential advantages and enhancements that QGNNs offer in contrast to classical GNNs, specifically concerning their ability to identify and mitigate fraudulent activities within intricate networks of financial transactions. By meticulously evaluating and comparing the performance, adaptability, and overall effectiveness of QGNNs benchmarked against classical GNNs, we intend to shed light on the transformative possibilities of leveraging QC for refining fraud detection methodologies.

Through this research endeavour, we aspire to comprehensively understand QGNNs potential to enhance the landscape of fraud detection strategies \cite{AS10}. By closely examining the strengths and constraints of QGNNs \emph{vis-\'{a}-vis} classical GNNs, our study seeks to contribute meaningful insights to the QC and fraud detection disciplines.

The structure of this paper is as follows:
\begin{itemize}
    \item Sec. \ref{sec:level2}, we present a comprehensive literature review on financial fraud detection, graph neural networks, and QML. We highlight the fundamental concepts of GNNs and emphasise the limitations of conventional methods.
    \item In Sec. \ref{sec:level3}, we examine the theory of GNNs and different models, focusing on \texttt{GraphSAGE} architecture.
    \item In Sec. \ref{sec:level4}, we conduct a comprehensive study of our proposed quantum graph neural network model for financial fraud detection.
    
    \item In Sec. \ref{sec:level5}, we provide a comprehensive understanding of our selected dataset characteristics, including attribute highlighting, class imbalance, and preprocessing steps.
    \item In Sec. \ref{sec:level6}, we evaluate the performance of QGNN and GNN on a real-world financial fraud dataset and discuss the results of our experiments.
    \item In Sec. \ref{sec:level7}, we summarise the key findings of our study and suggest directions for future research. 
\end{itemize}

\section{\label{sec:level2}Literature Review}

Transitioning our inquiry from the domain of conventional fraud detection techniques, we direct our attention towards the emerging promise of QGNNs. This transition represents a notable departure from established methodologies and carries the potential to reshape our understanding and implementation of fraud detection strategies. In adopting this advanced technological paradigm, we aim to uncover novel insights and methodologies that may prove instrumental in addressing the complexities of contemporary fraud detection challenges.

In the classical computing area, there exists a plethora of noteworthy research. We highlight some of the outstanding application-based research works below.

Ma \textit{et al.}, $2020$ \cite{AS20} provide an in-depth exploration of using Deep Learning (DL) techniques to detect anomalies in graph data. It emphasises the importance of anomaly detection in domains such as security, finance, and medicine. The paper highlights how conventional methods struggle with graph data complexities, leading to the rise of DL approaches. The authors systematically categorise and analyse contemporary techniques for graph anomaly detection based on their ability to identify different types of anomalous graph objects. The survey underscores the challenges inherent in this field and identifies future research directions. 

In today's rapidly evolving technological landscape, we are witnessing a growing variety of ways in which fraudulent transactions take place. This makes it crucial to understand how we detect fraud and to determine which approach works best to get accurate results. Researchers are on a widespread quest to explore different QML models that fit well for spotting fraud.

In \cite{AS25}, Ahmad \textit{et al.}, $2021$ explore the combination of QC and Deep Neural Networks (DNNs) to address network security through anomaly detection. Emphasising the transformative potential of QC, it develops a DNN-based quantum autoencoder to process quantum-encoded data. Applying dimensionality reduction techniques, the research enhances the model's ability to detect subtle anomalies within network data. The results reveal impressive accuracy rates of $100\%$ during training and $99\%$ during testing, demonstrating the framework's efficacy in capturing patterns and its potential to revolutionize anomaly detection in network security. This innovative research establishes a symbiotic relationship between QC and DL, offering a novel avenue for cybersecurity advancements.

In \cite{AS26}, Zheng \textit{et al.}, $2021$ introduce a Quantum Graph Convolutional Neural Network (QGCNN) model to address the challenge of applying a QNN to non-Euclidean spatial data, particularly graph-structured data. The QGCNN model efficiently captures graph topology and node features using quantum parametric circuits, making graph-level classification tasks effective. The study bridges the gap in QML research by extending QNNs to handle irregular spatial structures in graph data. 
The proposed QGCNN model showcases the potential of QC in enhancing graph-based ML tasks, contributing to the advancement of QML research.

QGNNs offer a transformative approach to fraud detection in the financial sector, addressing the escalating complexity of fraudulent schemes. By leveraging their capacity for intricate pattern recognition, enhanced data representation, and adaptability to evolving strategies, these networks hold the potential to significantly enhance the accuracy of fraud detection while mitigating false positives.

The study by Kyriienko \textit{et al.}, $2022$ \cite{AS23} focuses on utilising QML to enhance the detection of internal fraud in the financial sector, particularly fraudulent activities conducted by individuals within an organisation. The study employs quantum kernels generated through quantum feature maps based on Instantaneous Quantum Polynomial (IQP) circuits to transform classical data from credit card transactions into a higher-dimensional space for analysis. Applying the one-class support vector machine (OC-SVM) algorithm within this quantum feature space enables to detection of abnormal transactions. Simulations confirm the effectiveness of this quantum kernel-based approach in detecting internal fraud, highlighting the potential of QC to improve fraud detection accuracy in complex financial data.
 
Recent research efforts, such as \cite{AS21}, Innan \textit{et al.}, $2023$ investigate the utility of QML models in enhancing fraud detection within the financial sector. The authors address the pressing need for accurate fraud detection and propose leveraging QML techniques to improve system efficiency. Through a comparative analysis of four QML models –- Quantum Support Vector Classifier, Variational Quantum Classifier, Estimator Quantum Neural Network, and Sampler Quantum Neural Network –- the study assesses their performances in detecting financial fraud. Employing a synthetic dataset derived from BankSim, the authors employ data preprocessing, logical analysis, and visualisations to identify key discriminatory features. The QSVC emerges as the most effective model, yielding high F1 scores, exceeding $98\%$ for both fraud and non-fraud classes. The study highlights the potential of QML in fraud detection, providing insights into its capabilities and implications while acknowledging areas for future advancement.

In \cite{AS19}, Xiang \textit{et al.}, $2023$ address the challenge of credit card fraud detection by introducing a novel semi-supervised approach utilising attribute-driven graph representation and a Gated Temporal Attention Network (GTAN). The proposed method effectively enhances fraud detection performance, especially with limited labelled data, according to the study's key findings. The authors constructed a temporal-transaction graph based on transaction records, incorporating temporal transactions and interactions, and utilised GTAN for message passing and learning transaction representations. Risk patterns are further modelled through risk propagation amongst transactions. The research showcases that GTAN outperforms state-of-the-art baselines in detecting fraud across multiple datasets, demonstrating its ability to perform well with only a small fraction of labelled data. The study contributes a comprehensive framework that amalgamates attribute-driven embeddings, temporal attention mechanisms, and risk propagation to advance fraud detection methodologies, offering promising avenues for future research and innovation in the field.

In \cite{AS22}, Schetakis \textit{et al.}, $2023$ explore the application of fraud detection for small and medium-sized enterprises. The study employs a hybrid approach, comparing quantum classifiers with classical Artificial Neural Networks (ANN), demonstrating more efficient training and potential advantages in credit scoring. The study focuses on a classical-quantum hybrid model's performance using quantum classifiers and classical ANNs, revealing significant efficiency improvements with promising outcomes in credit scoring for SMEs.

In \cite{AS24}, Beer \textit{et al.}, $2023$ delve into the realm of QML by investigating the use of quantum neural networks for processing graph-structured quantum data. Addressing challenges and opportunities in the Noisy Intermediate-Scale Quantum (NISQ) devices era, the study introduces a novel concept of associating quantum states with graph vertices and correlations with edges. The paper proposes using modified loss functions and quantum training algorithms to learn and characterise complex graph structures with QNNs. The research contributes to the taxonomy of QML approaches, emphasising the significance of correlations and graph structure in advancing quantum information processing capabilities for various applications.

\section{\label{sec:level3}Classical Graph Neural Networks}

In this section, we present the theoretical foundations of classical graph neural networks, focusing on the type used in this study. Firstly, we briefly describe the mechanics of ANNs and thereafter describe the GNNs.

Artificial Neural Networks are a class of ML algorithms whose structure is derived from the workings of the Biological neurons in sentient beings. Analogous to how the neurological systems of these beings transmit information from one neuron to another, the ANN consists of different layers of neurons that transmit information. Below, we clarify further.
\begin{itemize} 
\item \textbf{Neurons:}
These are the basic units of computation in the ANN. Each neuron consists of an input, hidden, and output layer.
\begin{itemize} 
\item The input layers ingest the initial data. 
\item The hidden layer transforms the data received by the input layer and performs certain computations.
\item The output layer then receives the transformed data from the hidden layer and returns the predictive result.
\end{itemize}

\item \textbf{Weights and Biases:} 	 
Each connection between neurons has an associated weight, $w_{ij}\in\left[0,1\right]$, between nodes $i$ and $j$ respectively, which indicates the strength of connections between neurons and is learned and updated during the training process. The actual interpretation of weights is different for different types of ANNs. 

In an ANN, a bias term is added to the weighted sum of the inputs to a neuron before the activation function is applied. The bias term allows the neuron to learn a non-linear decision boundary, which is necessary for ANNs to learn complex functions.

\begin{figure}[H] 
    \centering
    \includegraphics[width=1\linewidth]{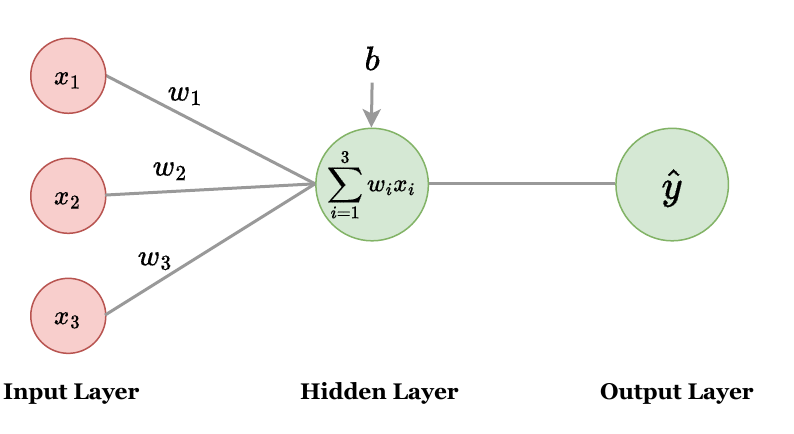}
    \caption{The basic structure of an ANN shows the weights and bias along with the respective layers.}
    \label{fig:fig1}
\end{figure}
In Fig.~\ref{fig:fig1}, the inputs are represented by $x_1, x_2, x_3$, and the weights are represented by $w_1,w_2,w_3$ shown in a summation form in the second layer.

In terms of the bias, $b$, the output layer is expressed as
\begin{align}
\hat{y} = \sum_{i} w_i x_i + b = w_{1}x_{1}+w_{2}x_{2}+w_{3}x_{3}+b.
\end{align}

\item \textbf{Activation Functions:} 
The ANNs consist of an activation function that transforms the given output non-linearly. 
Several examples of activation functions include the sigmoid, the ReLU and its variations, the hyperbolic tangent, the swish, etc., as shown in Fig.~\ref{fig:fig2}. The choice of each of these functions is an artefact of the problem at hand that one is trying to solve, and each has its advantages and disadvantages.

\begin{figure}[H]
    \centering
    \includegraphics[width=1\linewidth]{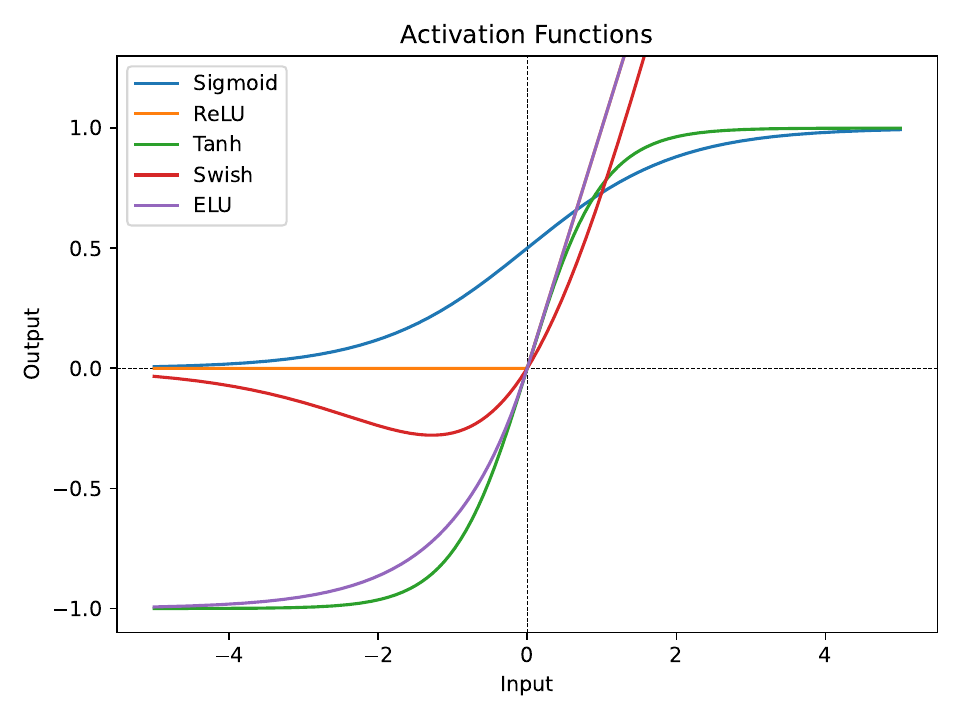}
    \caption{Plot showing the ReLU, Sigmoid, and Tanh activation functions along with vertical and horizontal lines at \(x=0\) and \(y=0\), respectively.}
    \label{fig:fig2}
\end{figure}

\item \textbf{Final Output:} The final layer of a basic ANN is the output layer, which yields the output of the computations from the hidden layers. 
\end{itemize}
As we discussed, artificial neural networks are ML models that can be used to learn from data. However, ANNs cannot capture the relationships between entities in a sequence or a grid, as they cannot consider the complex relationships between entities. To solve this problem, we use the where graph neural networks.

GNNs are a type of ANN specifically designed to process data structured as a graph. A graph is a group of nodes and edges representing relationships between them. The nodes in a GNN can represent entities such as people, objects, or places, and the edges can represent relationships such as friendships, kinship, or spatial proximity.

GNNs can process graph-structured data by iteratively aggregating information from neighbouring nodes. This allows GNNs to learn the complex relationships between entities in a graph, making them a powerful tool for node classification, link prediction, and graph classification tasks.

To express the edges and nodes of a graph as inputs to a neuron in an ANN, we first construct an adjacency matrix $\mathbf{A}$. It is, essentially, a matrix of dimensions $n\times n$ where $n$ is the number of nodes in a graph. The entries in the matrix are $1$ if an edge exists between the nodes and $0$ if there is no edge. Mathematically for each $A_{ij}\in\mathbf{A}$, and $1\leq i,j\leq n$, we have that
\begin{equation}
A_{ij}=
\begin{cases}
1,\text{if there exists an edge between nodes}\;i\;\text{and}\;j, \\
0,\text{otherwise}.
\end{cases}
\end{equation}
The adjacency matrix approach has several advantages: If the graph is undirected, the adjacency matrix becomes symmetric and has several interesting properties, including the diagonals representing self-loop vertices, the degree of the graph can be inferred from the sum of any row or column, the number of edges in the graphs (the number of connections required in the ANN) is just half the count of the number of nonzero entries, and properties relating the eigenvalues of $\mathbf{A}$.

\begin{figure}[H]
    \centering
    \includegraphics[width=0.7\linewidth]{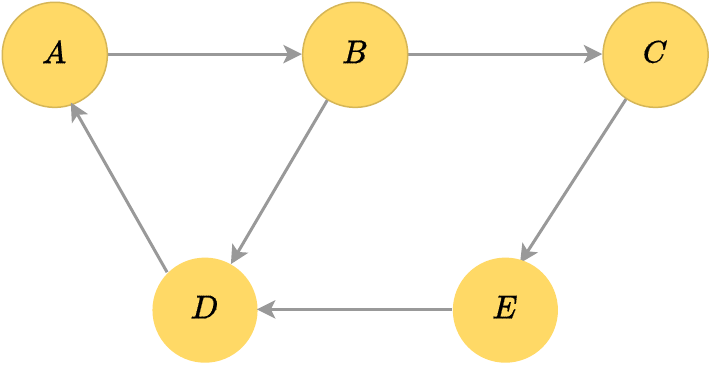}
    \caption{A directed graph. The arrows indicate the direction of flow. The graph consists of nodes labelled \(A\), \(B\), \(C\), \(D\), and \(E\).}
    \label{fig:fig3}
\end{figure}

In Tab.~\ref{tab:adjacency}, we observe no self-loops in the graph structure; thus, the diagonal entries of the adjacency matrix will be $0$,  This is illustrated in Fig.~\ref{fig:fig3}, which shows a directed graph with no self-loops.

\begin{table}[H] 
\caption{Adjacency matrix representing directed edges between nodes \(A\), \(B\), \(C\), \(D\), and \(E\).}
\centering
\begin{tabular}{|c|c|c|c|c|c|}
\cline{2-6}
\multicolumn{1}{c|}{} & A & B & C & D & E \\
\hline
A & 0 & 1 & 0 & 0 & 0 \\
\hline
B & 0 & 0 & 1 & 1 & 0 \\
\hline
C & 0 & 0 & 0 & 0 & 1 \\
\hline
D & 1 & 0 & 0 & 0 & 0 \\
\hline
E & 0 & 0 & 0 & 1 & 0 \\
\hline
\end{tabular}

\label{tab:adjacency}
\end{table}

Below, we elaborate on the critical components of GNNs.
\begin{itemize}

\item \textbf{Graph Convolution}
This is the key concept underpinning GNN architectures. Graph convolution predicts the features of the node in the next layer as a function of the neighbours’ features. It transforms the node’s features $x_{i}$ in a latent space $h_{i}$ that can be used for various reasons, as shown in Fig.~\ref{fig:fig4}.
\begin{figure}
    \centering
    \includegraphics[width=1\linewidth]{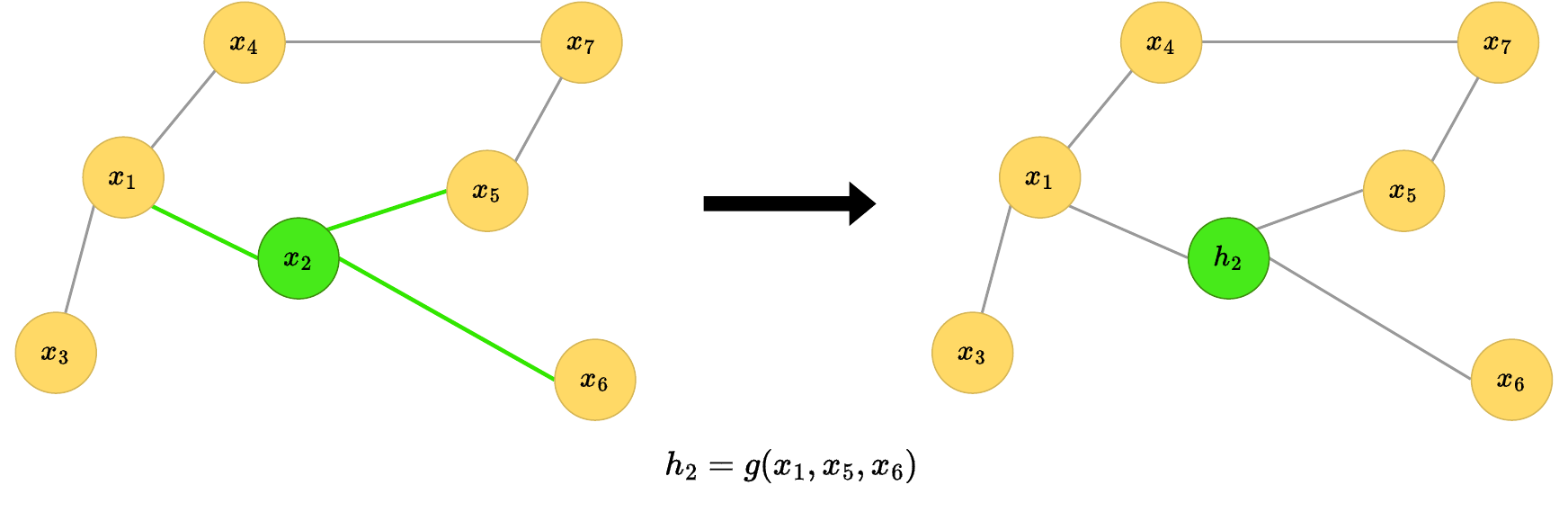}
    \caption{A graphical representation showing the dependency of the feature of nodes on the features of its neighbouring nodes.}
    \label{fig:fig4}
\end{figure}

\item \textbf{Node classification}
For a GNN, the objective is firstly to generate latent, from which we would then be able to work on a wide variety of standard tasks. Mathematically, we have that 
\begin{equation}
Z_{i}=f(h_i),
\end{equation}
where $h_{i}$ is the feature vector associated with node $i$, and it can also be represented in the Fig.~\ref{fig:fig5}.

\begin{figure}
    \centering
    \includegraphics[width=1\linewidth]{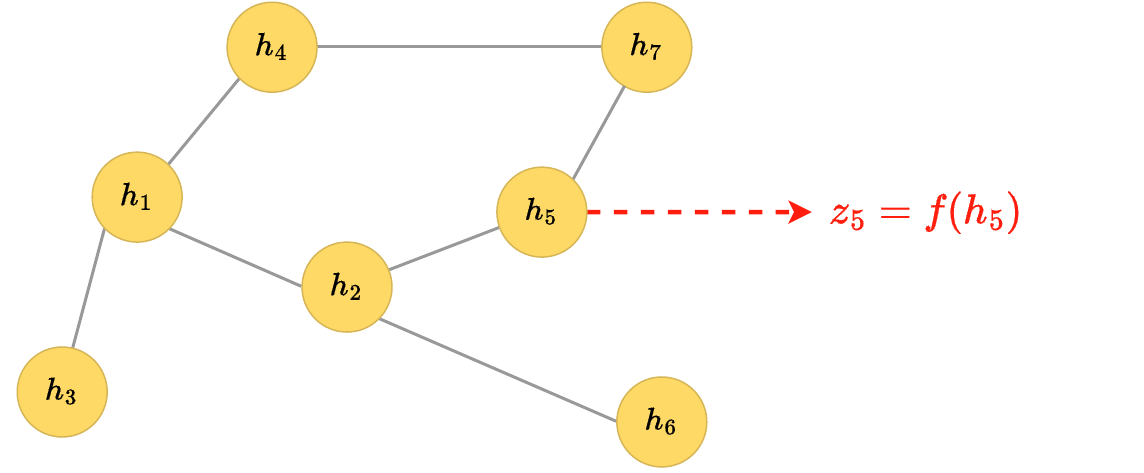}
    \caption{Graphical representation illustrating node classification. Nodes are associated with expressions \(Z_i = f(h_i)\), where \(i\) ranges from 1 to 7, capturing relationships between features \(Z_i\) and hidden features \(h_i\).}
    \label{fig:fig5}
\end{figure}

\item \textbf{Edge Classification:}
Similarly, we can use it to classify edges based on their features. To accomplish this, we generally need both the adjacent node vectors and the edge features, if they exist. Mathematically, we have that
\begin{equation}
Z_{ij}=f(h_i,h_j,e_{ij}),    
\end{equation}	
where $h_{i}, h_{j}$ are the feature vectors associated with nodes $i$ and $j$ respectively, and $e_{ij}$ is the edge connecting nodes $i$ and $j$ as shown in Fig.~\ref{fig:fig6}.

\begin{figure}
    \centering
    \includegraphics[width=1\linewidth]{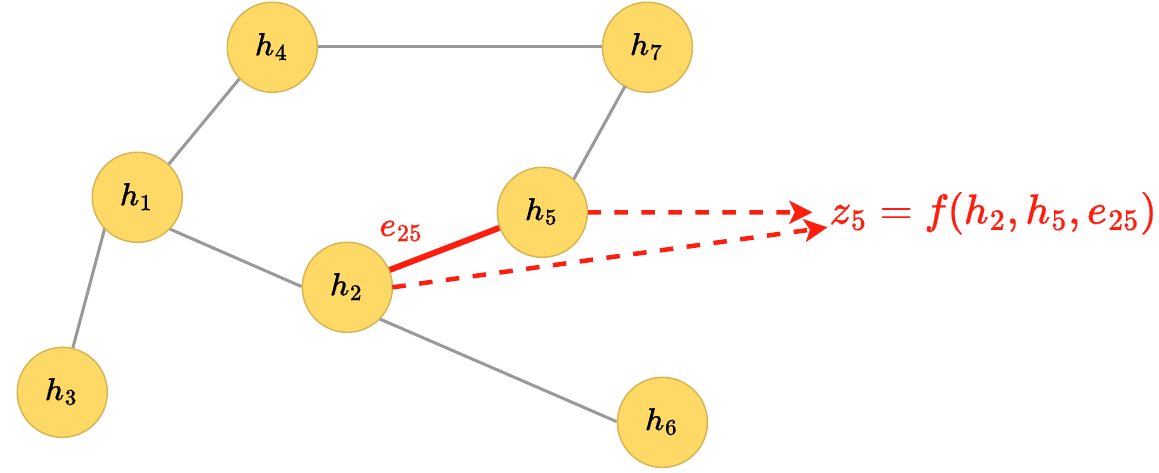}
    \caption{Graphical representation showcasing edge classification}
    \label{fig:fig6}
\end{figure}
\end{itemize}
\subsection{Types of Graph Neural Networks}

There are a variety of graph neural networks. Below, we focus on some of the important types of GNNs:

\subsubsection{Feed-forward GNNs:}

\begin{figure}
    \centering
    \includegraphics[width=1\linewidth]{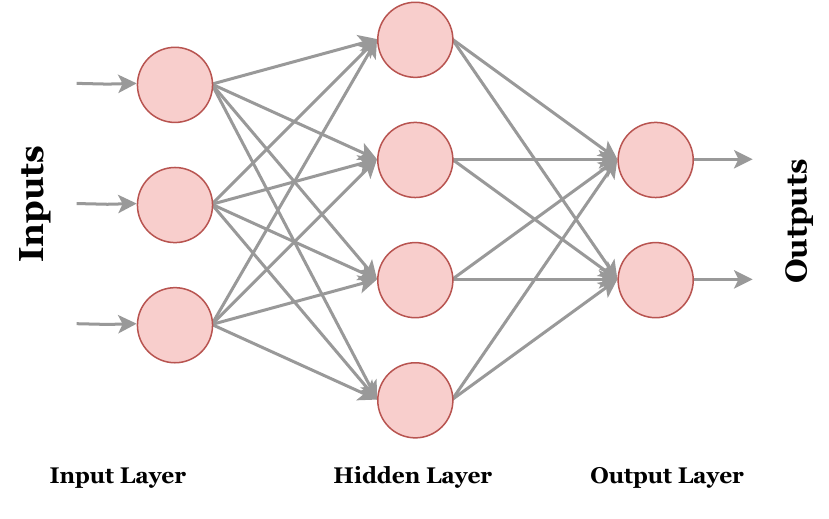}
    \caption{Basic structure of Feed-Forward GNNs. The network consists of an input layer with $3$ nodes, a hidden layer with $4$ nodes, and an output layer.}
    \label{fig:fig7}
\end{figure}

Also known as ``Deep feedforward Networks'' or ``Multi-layer Perceptrons''. In these model architectures, input data is propagated through a graph of neurons to produce output by applying the transfer function at the edge weights between each node as represented in Fig.~\ref{fig:fig7}. Feedforward graph neural networks are simplified versions of GNNs that work straightforwardly. They process graph data in just one pass and update node information using neighbouring node data.

\begin{figure*}[htpb]
    \centering
    \includegraphics[width=1\linewidth]{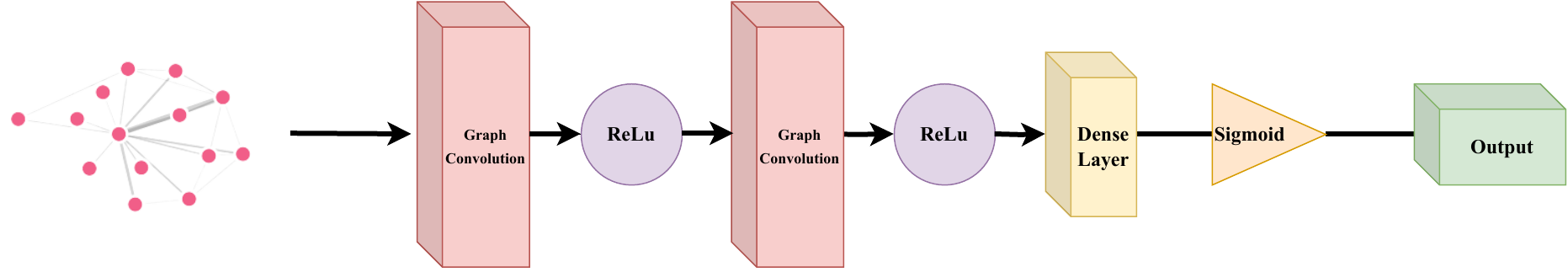}
\caption{Architecture of Graph Convolutional Neural Network. The network includes a convolution layer followed by a ReLU activation function, a subsequent convolution layer with another ReLU activation function, an intermediate dense layer, and finally, a sigmoid activation function before the output layer.}
    \label{fig:fig8}
\end{figure*}  
\subsubsection{Graph Recurrent Neural Networks (GRNNs):} 
\begin{figure}
    \centering
    \includegraphics[width=1\linewidth]{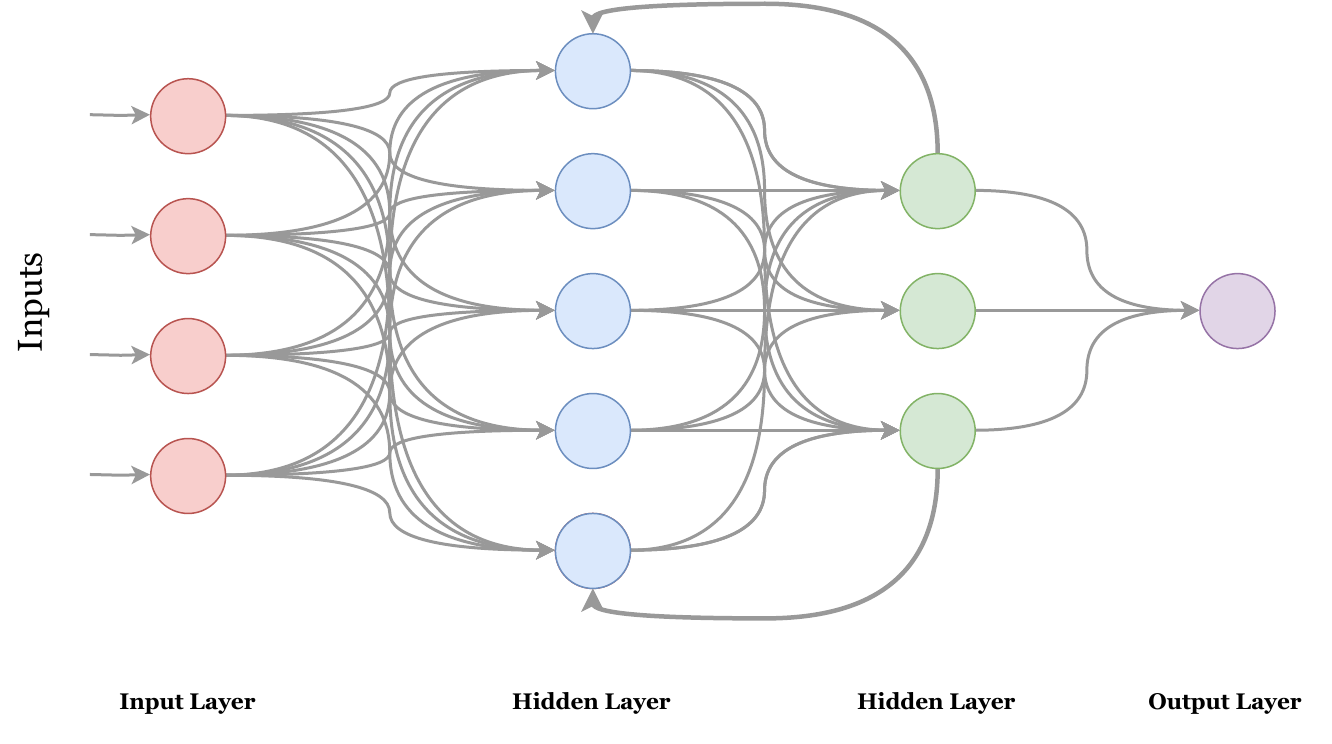}
    \caption{Structure of a Graph Recurrent Neural Network. The GRNN consists of an input layer with $4$ nodes, a hidden layer with $5$ nodes, and an additional layer with $3$ nodes. The architecture is completed with an output layer.}
    \label{fig:fig9}
\end{figure}

GRNNs relay information from previous to neighbouring nodes.
    At each step, the GRNN uses a hidden state as a memory of what the network has encountered thus far in the sequence, as represented in Fig.~\ref{fig:fig9}. The network is updated with each new step based on the current input and the previous hidden state. When working with sequences, the RNN ``unrolls'' itself through time, treating each step as a separate layer of the network. This helps the GRNN handle sequences of different lengths.
    In \cite{N4}, it was observed that Recurrent neural networks allow for loops and cycles, giving rise to dynamical systems and flexible behaviour in computation.

    In \cite{N5}, it has been observed that graph neural Networks are an evolution of Recurrent Neural Networks (RNNs) that address the limitations of RNNs by combining gated RNN architectures with graph convolutions.
  
\subsubsection{Graph Convolutional Neural Networks (GCNNs):} These architectures are one of the most popular and widely used GNNs in the literature, and they can be represented as shown in Fig.~\ref{fig:fig8}. The main property of GCNNs is their ability to maintain stability under stochastic perturbations. This means that GCNNs are less likely to be affected by noise or randomness in the data, which can be a major advantage in many applications. In \cite{N1}, a new RES and a GFT model were employed over a GCNNs to test the robustness of the graph filter. GCNN is stable to stochastic perturbations with a factor proportional to the link loss probability. In particular, a wider and deeper GCNN decreases the stability while improving the performance, indicating a trade-off between these two factors. In \cite{N2}, the study leverages the stability property of GNNs to seek stable representations within a distribution.

\subsubsection{\texttt{GraphSAGE} (Graph Sample and Aggregated) Networks:}

\texttt{GraphSAGE} is a node classification algorithm that samples and aggregates information from a node's local neighbourhood in a graph.

It improves over GCNNs by allowing nodes to aggregate information from a sampled set of neighbours, making \texttt{GraphSAGE} more scalable and generalisable to different graph structures.  

In \cite{N6}, Lo and \textit{et al.} proposed \texttt{E-GraphSAGE}, a GNN-based intrusion detection system which outperforms state-of-the-art methods in evaluating network-based intrusion detection systems.

These networks use a message-passing scheme to update node representations by aggregating and transforming information from the neighbourhood. They have been shown to be effective in capturing both the edge features and topological information of a graph, making them suitable for analysing interconnected patterns in network flows. The modified \texttt{E-GraphSAGE} and \texttt{E-ResGAT} algorithms integrate residual learning into GNNs to address the challenge of class imbalance and improve performance in minority classes.

\begin{figure*}[htpb]
    \centering
    \includegraphics[width=1\linewidth]{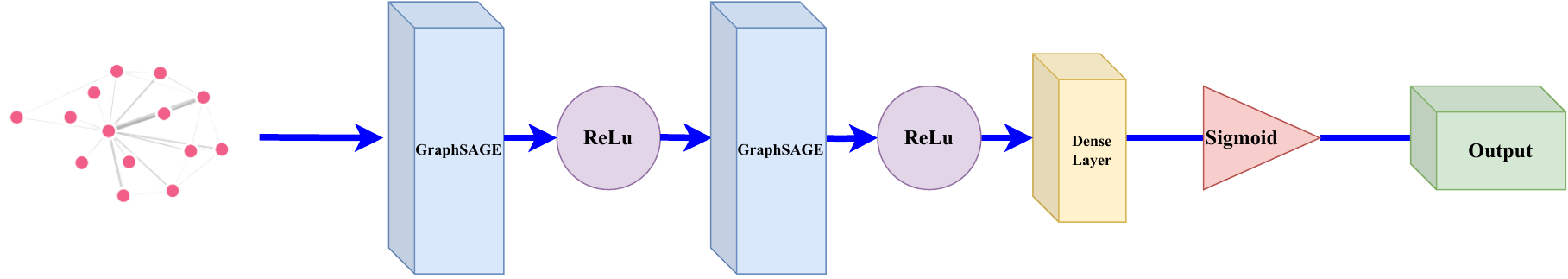}
  \caption{Architecture of the GraphSage. It comprises successive \texttt{GraphSAGE} layers, each followed by a ReLU activation function. An intermediate dense layer bridges the structure, culminating in a sigmoid activation function in the output layer.}
    \label{fig:fig10}
\end{figure*}

The \texttt{GraphSAGE} model architecture includes two graph convolution layers and uses the mean aggregation method to effectively capture node information from neighbours in the graph, as shown in Fig.~\ref{fig:fig10}.

For homogeneous graphs, the feature update rule is the mean aggregator, where the aggregation of features of the neighbours of the node \textit{v} are given by
\begin{equation}
    h_{N(v)}^{i} = \frac{1}{||N(v)||}D_{p}\left[h_{u}^{i-1}\right],\forall u \in N(v), \label{1}
\end{equation}
where the forward passes through layer \textit{k}  can be added in the future nodes and is given by
\begin{equation}
    h_{v}^{i} = \sigma\left(\text{concat}\left[W^{k}_{\text{self}}D_{p}\left[h_{v}^{i-1}\right],W^{i}_{\text{neigh}}h_{N(v)}^{i}\right]+b^{k}\right).\label{2}
\end{equation}
In Eq.\eqref{1} $D_p\left[\ldots\right]$ is a random dropout with probability $p$ applied to its argument vector, and $b^{k}$ is an optional bias.

Eqs.\eqref{1} and \eqref{2} are the rules deciding how the features get updated in the GNN. The \texttt{GraphSAGE} algorithm was implemented on the dataset to perform predictions using \texttt{StellarGraph}. From the original graph, $30\%$ of edges were removed to form a test set, $65\%$ acted as the train set, and $5\%$ acted as the validation set where the internal parameters were optimised over $10$ epochs using the ADAM optimiser. The classifier was then applied to the test set; Scores were generated for the classifier’s predictions and are provided in the Sec. \ref{sec:level6}. In the following, we describe the steps of the \texttt{GraphSAGE} algorithm.

\begin{algorithm}
\caption{\texttt{GraphSAGE}}
\label{GraphSAGE}
\begin{algorithmic}
\REQUIRE data (training, validation)
\FORALL{data in training, validation}
    \STATE Assign unique node identifier to sample
\ENDFOR
\STATE Create a \texttt{StellarGraph} $G$ using data with source column as ``node\_id''
\STATE Initialise a \texttt{GraphSAGENodeGenerator} with graph $G$, batch size $5$, and number of samples $\left[2, 32\right]$
\STATE Extract node labels $y$ from data
\STATE Generate a flow using the \texttt{GraphSAGENodeGenerator} and $y$ as targets

\IF{data is training}
    \STATE Initialise a \texttt{GraphSAGE} model with layer sizes $\left[128, 128\right]$ and activations [``relu'', ``relu'']
    \STATE Define input tensor and output tensor using the \texttt{GraphSAGE} model
    \STATE Add a Dense layer with $1$ unit and ``sigmoid'' activation to obtain predictions
    \STATE Create a \texttt{Keras} model with input and output tensors
    \STATE Compile the model with ``adam'' optimiser, ``binary\_crossentropy'' loss, and ``accuracy'' metric
    
    \FOR{epoch in $1$ to $10$}
        \STATE Train the model using the flow and perform validation for $1$ epoch
    \ENDFOR
    \ENDIF
\IF{data is validation}
    \STATE Predict on the flow using the trained model.
    \STATE Calculate mode threshold and convert predictions to binary
    \STATE Calculate precision-recall curve points and AUC-PR using true labels and predicted scores
\ENDIF
\end{algorithmic}
\end{algorithm}

The algorithm \ref{GraphSAGE} first assigns unique node identifiers to each sample in the dataset. A graph is then constructed using the \texttt{StellarGraph} library, where nodes are identified by their unique identifiers. The \texttt{GraphSAGENodeGenerator} is then used to generate a stream of node samples and their corresponding labels.
During the training phase, the \texttt{GraphSAGE} model is instantiated by specifying the sizes of the layers and activation functions. The model includes a dense output layer with a sigmoid activation function for generating predictions. After defining the input and output tensors, a \texttt{Keras} model is created and compiled using the Adam optimiser, binary cross-entropy loss, and accuracy metric. The model is trained over a number of epochs using the generated data stream.

For non-training data, the trained model is used to predict labels. The predictions are then post-processed by calculating the mode threshold and converting the predicted scores to binary form. Precision-recall curves are then computed to evaluate the model's performance, and the area under the precision-recall curve (AUC-PR) is used to quantify its effectiveness.

\section{\label{sec:level4}Quantum Graph Neural Networks} 
Our research methodology draws inspiration from the pioneering work of Hu \textit{et al.}, $2022$ \cite{Q3} in quantum graph neural network development. Our approach unfolds in the following manner:
\begin{itemize}

\item \textbf{Step $1$: Graph Construction and Representation}

Our initial step involves constructing the graph, where each graph corresponds to dataset transactions. This dataset encompasses $30$ encoded attributes, predominantly processed via Principal Component Analysis (PCA) techniques, except for two attributes – transaction amount and time. In this framework, nodes represent encoded features. At the same time, edges symbolise their fusion to get novel graph configurations, and each graph in our feature grouping employs interconnected nodes, totalling $28$ nodes per graph.
\item \textbf{Step $2$: Topological Data Analysis (TDA) for Graph Projection}

Utilising topological data analysis, the graph is embedded into a one-dimensional scatter plot. This technique discriminates between graphs based on their distinct elements, facilitating differentiation between fraudulent and non-fraudulent cases. Following scatter plot creation, node clustering reduces nodes while encapsulating the initial $28$ features, minimising quantum resource requirements. A final $28$-dimensional feature vector representing the positional attributes is generated.
\item \textbf{Step $3$: Encoding within Quantum Circuit and Linear Transformation}

Upon acquiring the graphical representation of each transaction, the subsequent step entails encoding it into a quantum state utilising the angle-encoding technique, employing a range of qubits from $1$ to $28$. To initiate, the encoding process is applied to all node-level features ($N \times 28$), resulting in a dimensionally apt vector ($N \times q$) facilitated by a fully connected linear layer. This vector is subsequently channelled into the designated quantum circuit with $q$ qubits as defined

\begin{equation}
    \ket{x} = \bigotimes_{i=1}^{N}\cos(x_{i})\ket{0} + i\sin(x_{i})\ket{1}.
\end{equation}

\item \textbf{Step $4$: Enhancing Node Features with Variational Quantum Circuits}

Following the previous steps, a sequence of two-local ansatz constructs orchestrates the implementation of a multi-layered Variational Quantum Circuit (VQC) calibrated for the transactional dataset. Fig.~\ref{fig:Q_circ} visually portrays a circuit encompassing a $6$-qubit with the angle encoding alongside $2$ concealed VQC layers. Notably, the ansatz capitalises on \texttt{Rx}, \texttt{Ry}, and \texttt{CNOT} operations while adopting \texttt{Rx} rotations for angle encoding, as explained in the $3$rd step. Subsequently, circuit measurement culminates in using Pauli-\texttt{Z} expectation. This strategic circuit design is a conscientious outcome of considerations concerning circuit expressibility and intricate node-node interactions embedded within graphical data.

\begin{figure*}[htpb]
    \centering
    \includegraphics[width=0.75\linewidth]
    {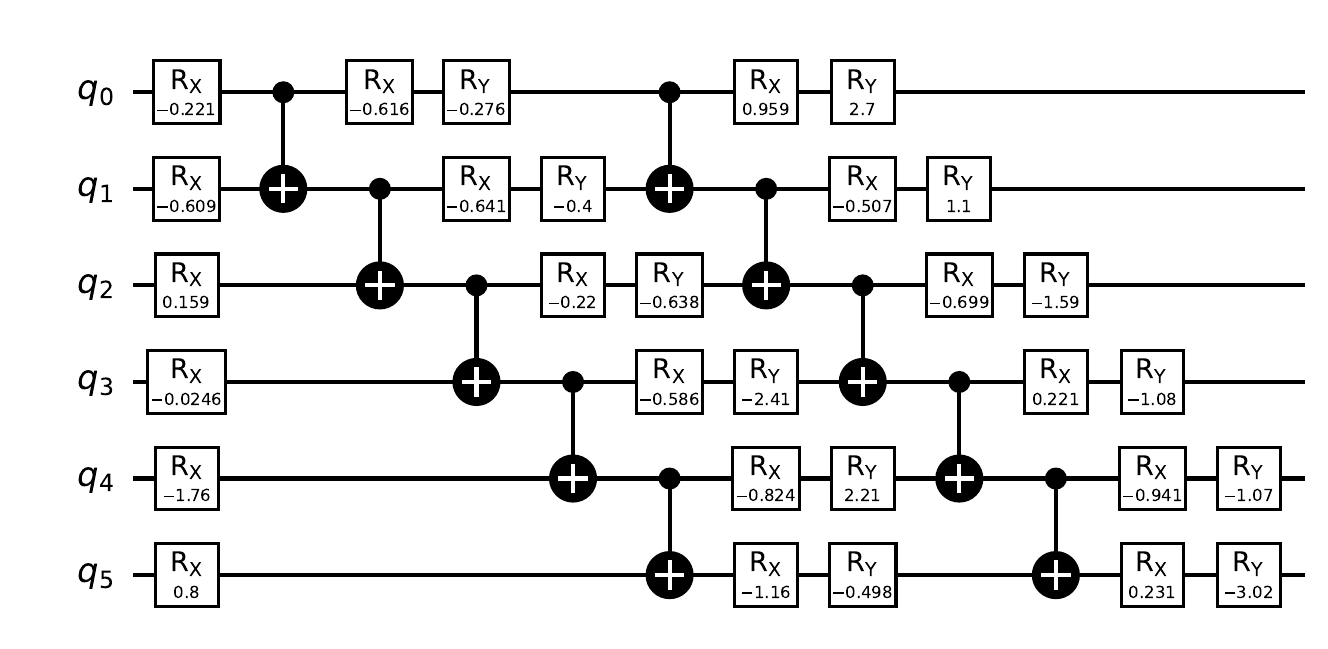}
    \caption{Quantum circuit for angle encoding and VQC in the QGNN architecture with RY, RX, and CNOT gates.}
    \label{fig:Q_circ}
\end{figure*}

We enhance node features using the VQC, emulating the operational principles of conventional classical graph neural networks with fully connected layers. VQC entails the successive application of variational layers. A single layer transformation $L$ is depicted as

\begin{equation}
    L: \ket{\psi(x)}\xrightarrow{}\ket{\psi(y)} = U(w)\ket{\psi(x)},
\end{equation}

where $U(w)$ represents the VQC with $N$ layers, defined as \begin{equation}
    U(w) = T\bigotimes_{k=1}^{N}R_y(w_k)\ket{\psi(x)}.
\end{equation}

\item \textbf{Step $5$: Integration of Unitary Gates and Classification}

In order to meet the requirements of a quantum graph neural network, it is necessary to have unitary gates that act on two states simultaneously while preserving the others. As we explained, rotational gates are essential to integrate the graph into the quantum circuit. Once these gates are in place, a linear layer can classify the graph as fraudulent or non-fraudulent. The comprehensive procedure outlined in Fig.~\ref{fig:qgnnarch} sheds light on the architecture and operational aspects of the QGNN, highlighting the distinct components and their produced interplay in fraud detection. Additionally, Algorithm \ref{alg:qgnnalg} provides a step-by-step algorithm description.
\end{itemize}

\begin{figure*}[htpb]
    \centering
    \includegraphics[width=1\linewidth]{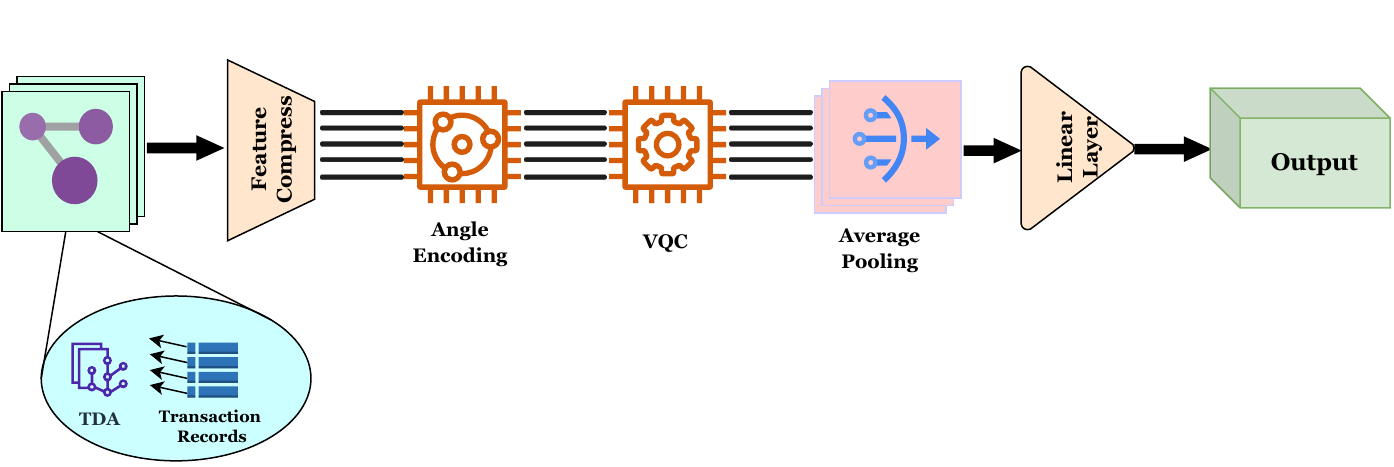}
    \caption{Architecture of the quantum graph neural network. The QGNN starts with a graph as input, followed by a feature compression step. Angle encoding prepares the data for the VQC processing. The output of the VQC is subjected to average pooling, then fed through a linear layer, and finally to the output layer}
    \label{fig:qgnnarch}
\end{figure*}

\begin{algorithm}
\label{alg:qgnnalg}
\caption{Quantum Graph Neural Networks}
\begin{algorithmic}

\setstretch{1} 
\STATE \textbf{Inputs:} Graph data (training, validation)
\STATE Initialise QGNN
\FOR{each epoch in training}
    \FOR{each graph batch in training data}
        \STATE Extract graph features and edges
        \STATE Apply angle encoding to the graph features
        \STATE Initialise Quantum Circuit for VQC
        \STATE Generate tunable parameters for VQC
        \STATE Update Quantum Circuit using inputs and VQC parameters
        \STATE Run the Quantum circuit (VQC) to obtain quantum output
        \STATE Perform average pooling for graph-level features

        \STATE Make predictions using a linear layer
        \STATE Compute loss and update parameters using Adam optimiser
    \ENDFOR
    \STATE Evaluate QGNN on validation data
\ENDFOR
\STATE Calculate optimal threshold for binary classification

\STATE Convert predictions to binary using the threshold
\STATE Calculate Precision-Recall curve and AUC-PR
\STATE \textbf{Return} Classification metrics, loss values
\end{algorithmic}
\end{algorithm}

\section{\label{sec:level5}The Dataset}

We used the publicly available credit card fraud detection dataset \cite{D1} for our analysis. This dataset contains credit card transactions of European cardholders with encrypted feature labels obtained from principal component analysis on actual user data. Each transaction is labelled as either fraud $(1)$ or not fraud $(0)$ and contains $28$ encrypted features along with amount and time values. The dataset is highly imbalanced, with only $492$ frauds out of $284\;807$ transactions. To address this imbalance, we used random undersampling on the non-fraud transactions.

\begin{figure}[htpb]
    \centering
    \includegraphics[width=1\linewidth]{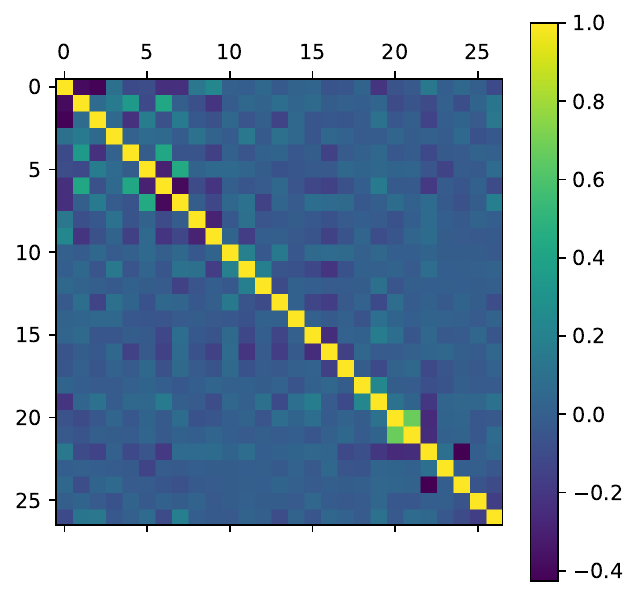}
    \caption{Correlation matrix \((v_1 - v_{28})\) of the dataset. The figure illustrates the relationships between the features \(v_1\) to \(v_{28}\) by displaying their correlation values.}
    \label{fig:Corr_mat}
\end{figure}

For a comprehensive understanding of the dataset characteristics, we utilised visual representations. In Fig.~\ref{fig:Corr_mat}, we present the correlation matrix $(v_1 - v_{28})$ of the dataset. The figure effectively illustrates the relationships between the features $v_1$ to $v_{28}$ by displaying their correlation values. This information offers insights into how these features interact and influence one another. 

\begin{figure*}[htpb]
    \centering
    \includegraphics[width=0.5\linewidth]{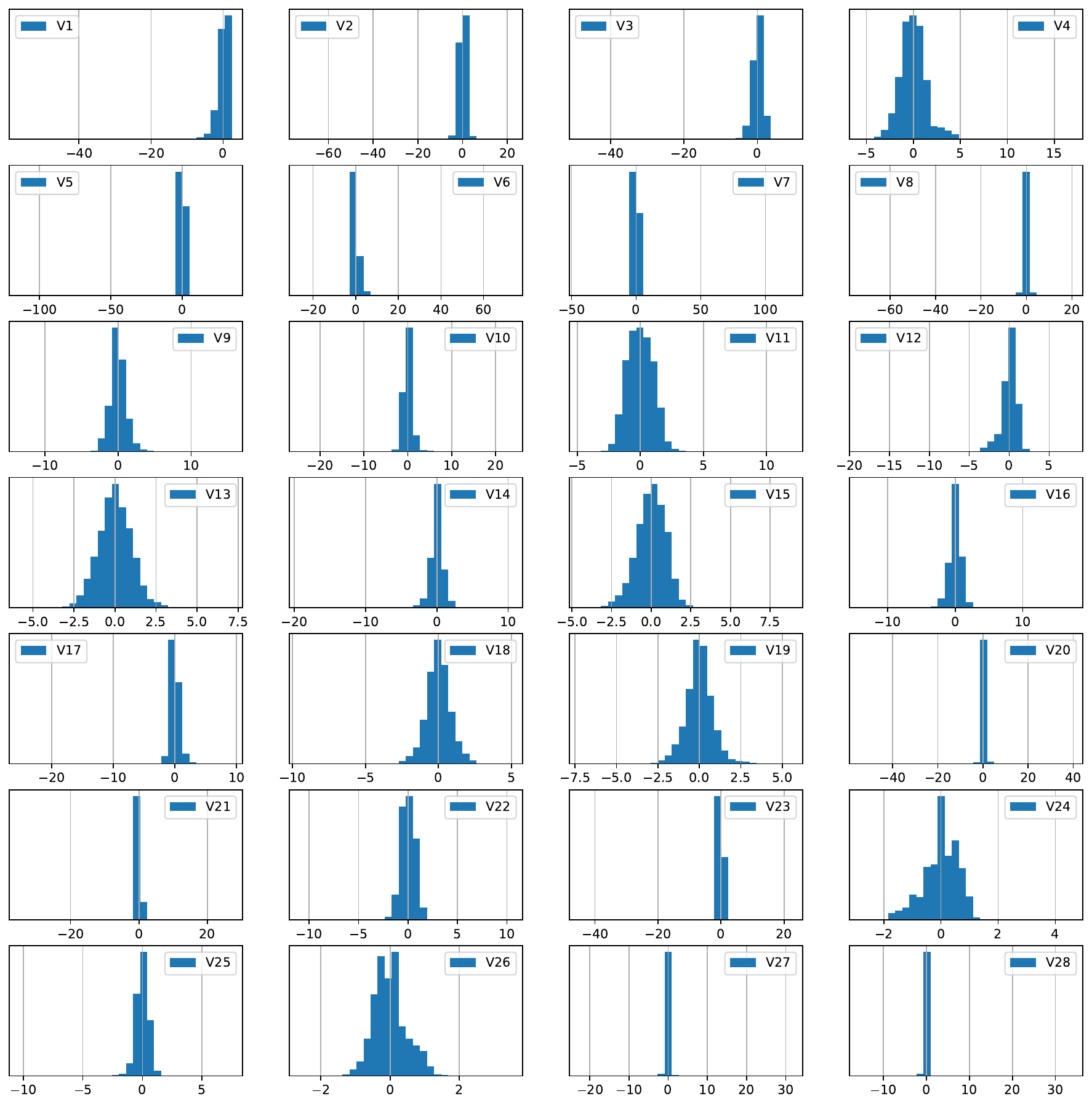}
    \caption{Histograms illustrating the distribution of all features in the dataset.}
    \label{fig:Dist_feat}
\end{figure*}

Additionally, Fig.~\ref{fig:Dist_feat} offers histograms that vividly depict the distribution of all features in the dataset. These histograms visually represent the data distribution patterns, assisting us in identifying potential trends and outliers. Furthermore, our approach involved translating each transaction as a graph using topological data analysis \cite{D2}. This technique allowed us to capture intricate relationships among the various feature vectors, enhancing our ability to analyse the data effectively. In the following steps, we explain how to apply and implement this technique.

Given a transaction described by a vector

\begin{equation}
\mathbf{V} = [\text{Time}, V_1, V_{2},\ldots,V_{28}, \text{Amount}],   
\end{equation}

We create the graph $G = \left(N,E\right)$, where $N\in\{R^{1}, R^{2},\ldots, R^{28}\}$ represents the feature vector, and $E$ represents the edge list, where, $V_{1}, V_{2}, \ldots, V_{28}$ denotes the encrypted feature labels in the dataset. 

Next, we focus on converting the transaction, $\left\{T_i\right\}_{i=1}^{284\;807}$, into its graphical representations $G_i$ via the following steps using TDA:
\begin{enumerate}
    \item Create a set $\{S_i\}$ of 3D data points, where $S_i  = \big\{P_{j} \; | \; P_{j} = \left[\text{Time}_i,V_j,\text{Amount}_i\right], \: \forall{j} \in [1,28] \: \big\} $. This helps in capturing the correlation of $V_j$ with time and amount. 
    \item Project all data points $P_{j} \in S_i$ onto a 1D plane. We denote these projected points as $f_{j}$.
    \item Draw covers around all the projected data points $\{f_{j}\}_{j=1}^{28}$ and cluster them using a suitable clustering algorithm, here, we have used DBSCAN. Let these clusters form a set $\{C_i\}$, where $C_{i} = \big\{ c_k \:|\: \exists_{j,k} \; : \; f_j \in c_k, \: \forall j \in [1,28] \; and \; 1 <  k < |C_i|  \big\}$
    \item Draw edges between clusters with common data points to form the graph $G_i$.
\end{enumerate}

Here, a node $N_{k} \in G_i $, represents a cluster $c_k \in \left\{C_i\right\}$, s.t. $N_{k} =\left(0\ldots V_j\ldots 0\right)_{28},\;\forall f_{j} \in c_k$. Thus, the relation between different features is captured and grouped using the clustering algorithm of high-dimensional data via a more representative graphical structure. Each graph represents a transaction labelled fraud or non-fraud for the classification task.

			
\section{\label{sec:level6}Results and Discussion}
We tested the performance of our two classical and quantum algorithms on the same dataset described in Sec. \ref{sec:level5}. The results showed that the \texttt{GraphSAGE} model achieved promising results on the fraud classification task. The model was trained for $10$ epochs using the Adam optimiser with a batch size of $5$ using Keras, as shown in Tab.~\ref{tab:hyperparameters}. The aggregation method was set to mean. These hyperparameters were chosen based on preliminary experiments and empirical observations to balance model complexity and generalisation performance.

\begin{table}[htpb]
    \centering
    \caption{Hyperparameters and training details for the \texttt{GraphSAGE} model.}
    \vspace{0.2cm}
    \begin{tabularx}{0.4\textwidth}{@{}lX@{}}
        \toprule
        \textbf{Hyperparameter} & \textbf{Value} \\
        \midrule
        Model Architecture & \texttt{GraphSAGE} \\
        Number of Epochs & $10$ \\
        Optimiser & Adam \\
        Batch Size & $5$ \\
        Aggregation Method & Mean \\
        \bottomrule
    \end{tabularx}
    \label{tab:hyperparameters}
\end{table}

\begin{figure}[htpb]
    \centering
    \includegraphics[width=1\linewidth]{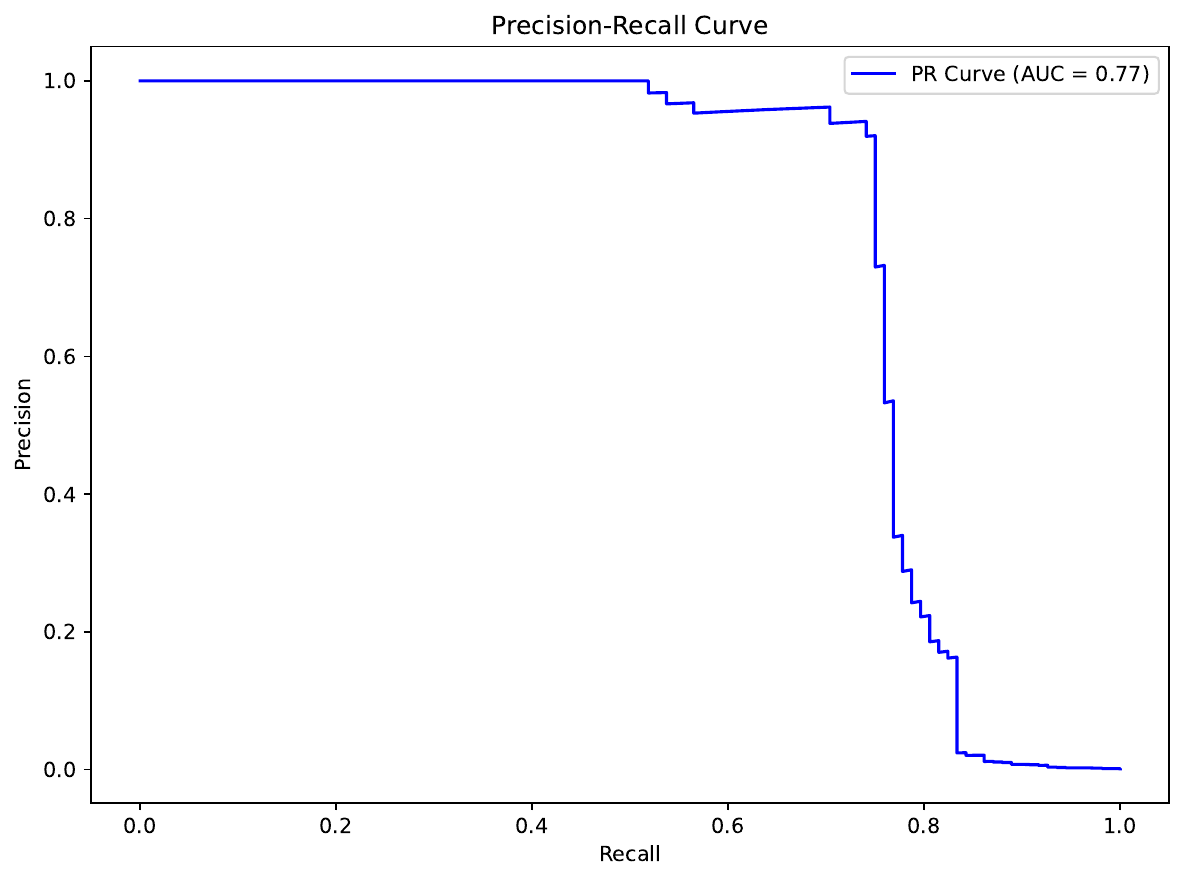}
    \caption{Receiver Operating Characteristic Curve for the \texttt{GraphSAGE} model with an AUC of $0.77$.}
    \label{fig:roc-auc}
\end{figure}

The receiver operating characteristic (ROC) curve presented in Fig.~\ref{fig:roc-auc} shows the trade-off between the true positive rate (TPR) and the false positive rate (FPR) for the \texttt{GraphSAGE} model. The Area Under Cover (AUC) score measures how well a model can distinguish between positive and negative instances. A higher AUC score indicates a better model. The \texttt{GraphSAGE} model achieves an AUC of $0.77$, within the acceptable range for fraud classification tasks. The performance metrics for the \texttt{GraphSAGE} model are also within the acceptable range, with an accuracy of $92.3\%$, and F1 score of  $0.83$. These scores are reported in Tab.~\ref{tab:Clas_res}. 

\begin{figure}[htpt]
    \centering
    \includegraphics[width=1\linewidth]{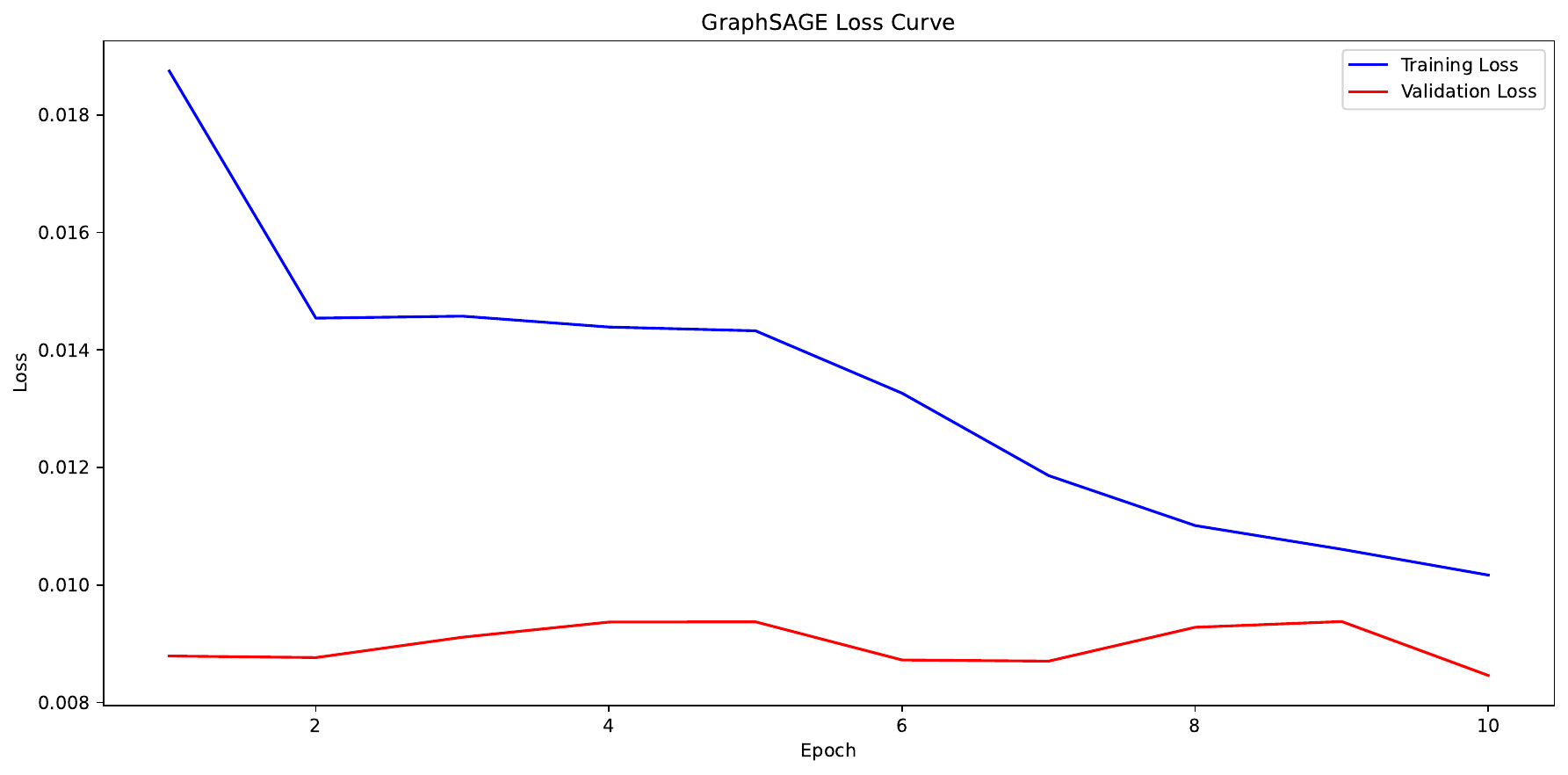}
    \caption{Training and validation loss curves for the \texttt{GraphSAGE} model.}
    \label{fig:loss-curves}
\end{figure}

\begin{table*}[htpb]
    \centering
    \caption{Performance test metrics of \texttt{GraphSAGE}}
    \vspace{0.2cm}
    \begin{tabular}{|c|c|c|c|}
        \hline
         \textbf{Accuracy (\%)} & \textbf{Precision (\%)} & \textbf{Recall (\%)} & \textbf{F1 score} \\
        \hline
        \hline
92.3 & 95.2 & 76.3 & 0.83 \\
        \hline
    \end{tabular}
    \label{tab:Clas_res}
\end{table*}
The loss curves in Fig.~\ref{fig:loss-curves} show that the training loss decreases steadily over time while the validation loss decreases more slowly, which suggests that the model is well-fitted on both the training and validation data.

We performed experiments on the quantum graph neural network using qiskit \cite{qiskit} under different scenarios, namely qubit count and hidden layers. The results in Tab.~\ref{tab:Q_res} show that a dense encoding scheme of node features using fewer qubits performed better than using more qubits and sparser representations, and this is evident as the model with $6$ qubits, with an accuracy of $94.5\%$ and F1 score of $0.86$, performed best compared to that with 16 qubits, with an accuracy of $92.0\%$ and F1-score of $0.81$.

\begin{table*}[htpb]
    \centering
    \caption{Performance test metrics of QGNN. The table presents results for various configurations, including the number of qubits and layers.}
    \vspace{0.2cm}
    \begin{tabular}{|c|c|c|c|c|c|}
        \hline
        \textbf{Num Qubits} & \textbf{Layers} & \textbf{Accuracy (\%)} & \textbf{Precision (\%)} & \textbf{Recall (\%)} & \textbf{F1 score} \\
        \hline
        \hline 
        6 & 1 & 94.5 & 96.1 & 79.5 & 0.86 \\
        16 & 1 & 92.06 & 86.5 & 77.3 & 0.81 \\
        6 & 2 & 91.5 & 93.3 & 65.6 & 0.76 \\
        16 & 2 & 89.2 & 86.6 & 72 & 0.77 \\
        \hline
    \end{tabular}
    \label{tab:Q_res}
\end{table*}

\begin{figure}[htpb]
    \centering
    \includegraphics[width=1\linewidth]{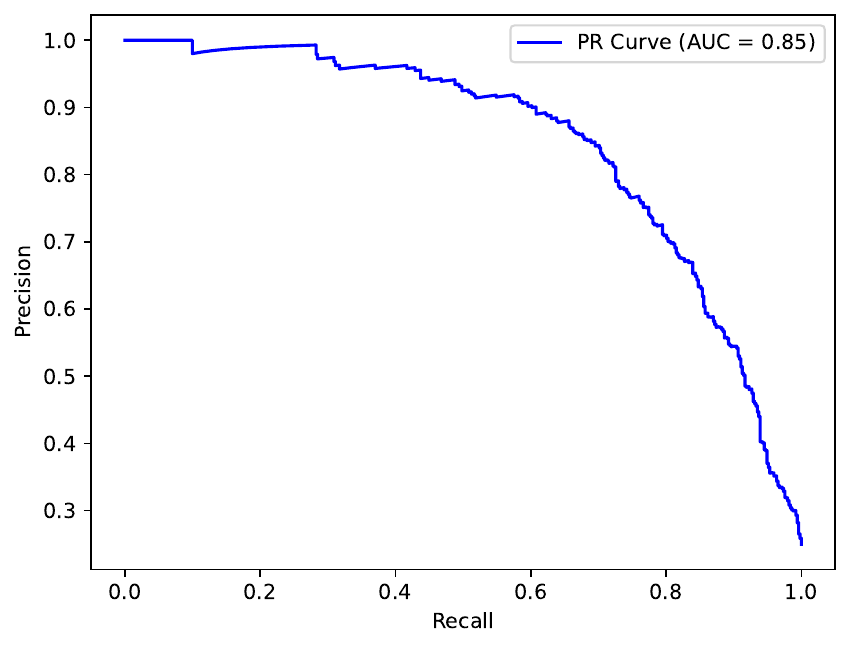}
    \caption{Receiver Operating Characteristic curve for QGNN model with $6$ qubits and $1$ hidden layer, with an AUC of $0.85$}
    \label{fig:roc-Q}
\end{figure}

\begin{figure}[htpb]
    \centering
    \includegraphics[width=1\linewidth]{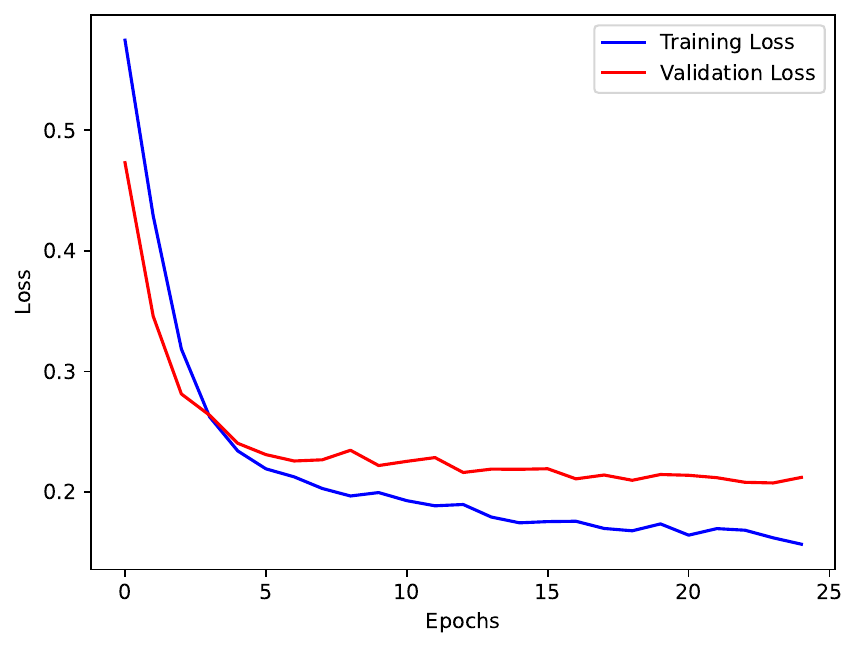}
    \caption{Training and validation loss curves for the QGNN model with $6$ qubits and $1$ hidden layer.}
    \label{fig:loss-q}
\end{figure}

The AUC for the precision-recall curve from Fig.~\ref{fig:roc-Q} is estimated to be roughly $0.85$, which shows high tolerance in our method towards biased datasets with high class imbalances, apart from having highly accurate predictions. A smoother transition curve leads to a better F1 score, and the loss curves show a decrease in the loss function over time for both training and validation, which shows that our model is well-fitted, as presented in Fig.~\ref{fig:loss-q}.

These results are observed on a basic network architecture of quantum circuits and outperform various classical approaches. This is due to the enhanced representation of graphical structures through quantum circuits and efficient feature relations using topological data analysis. We can see an improvement of $\approx3\%$ over classical GNN, such as \texttt{GraphSAGE}, across various metrics. However, it was observed that graphical techniques, in general, help enhance the precision and accuracy of these classification systems more than other non-graphical approaches.

Careful analysis of quantum techniques suggests that better representation, deeper search space, state entanglement, and qubit parallelism are key factors contributing to the supremacy of quantum circuits for classification tasks. In our experiments, we observed the same with a smaller system with 6 qubits and $\approx$200 parameters, which makes our model QGNN outperform the classical GNN.
These results suggest that with more advanced features and complex quantum systems, we can achieve a real quantum advantage in solving problems such as fraud detection in highly imbalanced datasets.


\section{\label{sec:level7}Conclusion}
Our research has demonstrated the potential of quantum graph neural networks in detecting financial fraud. We have compared the performance of QGNNs with classical graph neural networks and found that QGNNs outperformed GNNs in terms of accuracy and efficiency. Our experiments on a real-world financial dataset show that QGNNs can detect fraud with a remarkable accuracy of $94.5\%$, which is significantly higher than GNNs' accuracy of $92.3\%$. 

Moreover, we have also studied the influence of the number of qubits and layers on QGNNs' performance. Our research indicates that QGNNs with $6$ qubits and one layer exhibit the best performance, with an accuracy of $94.5\%$. This can be attributed to the fact that fewer qubits are less prone to noise, and a single layer is sufficient to grasp the essential features of the data. 

Our research shows that QGNNs have the potential to revolutionise fraud detection and improve the security of financial systems. It also highlights the need for further research to address the challenges of implementing QC in real-world applications.


			%
			%
\section*{Reference}
\bibliography{references}
		\end{document}